# Observation of superconductivity at 30~46K in $A_xFe_2Se_2$ (A=Li, Na, Ba, Sr, Ca, Yb, and Eu)

T. P. Ying[1], X. L. Chen[1,*], G. Wang[1,*], S. F. Jin[1], T. T. Zhou[1], X. F. Lai[1], H. Zhang[1], W. Y. Wang[1]

[1]Research & Development Center for Functional Crystals, Beijing National Laboratory for Condensed Matter Physics, Institute of Physics, Chinese Academy of Sciences, Beijing 100190, China

Correspondence and requests for materials should be addressed to X.L.C. (email: chenx29@iphy.ac.cn) or G.W. (email: gangwang@iphy.ac.cn).

New iron selenide superconductors by intercalating smaller-sized alkali metals (Li, Na) and alkaline earths using high-temperature routes have been pursued ever since the discovery of superconductivity at about 30 K in $KFe_2Se_2$, but all have failed so far. Here we demonstrate that a series of superconductors with enhanced $T_c$=30~46 K can be obtained by intercalating metals, Li, Na, Ba, Sr, Ca, Yb, and Eu in between FeSe layers by the ammonothermal method at room temperature. Analysis on their powder X-ray diffraction patterns reveals that all the main phases can be indexed based on body-centered tetragonal lattices with $a$~3.755-3.831 Å while $c$~15.99-20.54 Å. Resistivities show the corresponding sharp transitions at 45 K and 39 K for $NaFe_2Se_2$ and $Ba_{0.8}Fe_2Se_2$, respectively, confirming their bulk superconductivity. These findings provide a new starting point for studying the properties of these superconductors and an effective synthetic route for the exploration of new superconductors as well.

The discovery of superconductivity (SC) at about 30 K in $K_{0.8}Fe_2Se_2$[1], and subsequently in $Rb_{0.8}Fe_2Se_2$[2,3], $Cs_{0.8}Fe_2Se_2$[4,5], $(Tl,K)Fe_2Se_2$[6], and $(Tl,Rb)Fe_2Se_2$[7] has aroused a surge of research interests as these iron chalcogenides possess a strikingly distinct electronic structure from their iron pnictide counterpart superconductors[8-13]. Angle-resolved photoemission spectroscopy studies revealed that only electron Fermi surfaces are observed around the zone corners while no hole Fermi surface near the zone centre, implying that interband scattering or Fermi surface nesting is not a dominant pairing mechanism[14,15]. X-ray and neutron diffraction studies indicated that all these superconductors are close to $Am_{0.8}Fe_{1.6}Se_2$ (Am = K, Rb, Cs, and Tl) in composition with Fe vacancies ordering to form a $\sqrt{5}\times\sqrt{5}\times1$ supercell below the order-disorder transition temperatures[16-18], but the phase responsible for the SC is still in debate[19]. Moreover, a trace fraction of SC at 44 K was occasionally observed in an unknown phase[20] coexisting with $K_{0.8}Fe_2Se_2$. Efforts to synthesize new isostructural chalcogenides with other alkali metals and alkaline earths for clarifying the superconducting phase and exploring new superconductors have failed up to now.

Liquid ammonia (LA) is known to dissolve alkali metals, alkaline earths, and even some rare earths to varying extents. Intercalating alkali metals, alkaline earths, rare earth Yb, and ammonia into $MoS_2$[21-24] and $C_{60}$[25] in the LA yields 2.4~6.3 K and ~29 K superconductors, respectively. Here we report that metals, Li, Na, Ba, Sr, Ca, Yb, and Eu can be intercalated in between FeSe layers by the ammonothermal method and form a series of superconductors with enhanced $T_c$ = 30~46 K, which cannot be obtained by similar high-temperature routes described in ref. 1. These results demonstrate that metal intercalations at low temperatures are thermodynamically favored and SC can be induced by electron doping through intercalating a variety of metals. Our findings provide a new starting point for further exploring the SC and mechanism in the metal-doped iron selenides.

**Results**

Figure 1 shows the powder X-ray diffraction patterns for nominal $A_xFe_2Se_2$, (A = Li, Na, K, Ba, Sr, Ca, Yb, and Eu). The patterns for $\beta$-FeSe and $K_{0.8}Fe_{1.7}Se_2$ that were obtained by high-temperature route are also included for comparison. Trying to index each pattern based on a single crystallographic unit cell only succeeds for $LiFe_2Se_2$, $NaFe_2Se_2$, and $Ba_{0.8}Fe_2Se_2$, while patterns for $KFe_2Se_2$, $Ca_{0.5}Fe_2Se_2$, $Sr_{0.8}Fe_2Se_2$, $EuFe_2Se_2$, and $Yb_{0.8}Fe_2Se_2$ yield more than one crystallographic unit cell. The patterns for $LiFe_2Se_2$, $NaFe_2Se_2$, and $Ba_{0.8}Fe_2Se_2$ were indexed on body-centered tetragonal cells with lattice constants $a$ = 3.775(5) Å, $c$ = 17.04(3) Å; $a$ = 3.7846(4) Å, $c$ = 17.432 (1) Å; and $a$ = 3.7781(2) Å, $c$ = 16.8427(8) Å, respectively. All constants '$a$' are comparable with that of $K_{0.8}Fe_{1.7}Se_2$ by the high-temperature route while '$c$' being elongated. Since the much elongated unit cell along the $c$ axis, in between the FeSe layers should exist other atomic groups relating with $NH_3$ apart from the intercalated metals. Considering the relatively little contribution to the peak intensity from N and H and the heavy absorption of the Cu $K\alpha$ radiation by Fe, we first performed Rietveld refinements for nominal $NaFe_2Se_2$ based on the $I4/mmm$ structural model of $K_{0.8}Fe_2Se_2$[1] and the final agreement factors converged to $R_p$ = 2.18%, $R_{wp}$ = 3.05%, and $R_{exp}$ = 1.77%. Trying on the model proposed for $Li_x(NH_2)_y(NH_3)_{1-y}Fe_2Se_2$[26] does not improve the refinement, only leading to a slight increase in agreement factors. Based on $K_{0.8}Fe_2Se_2$ model, the final refined parameters are summarized in Table I. As shown in the inset of Fig. 2, the crystal structure obtained from the refinement is composed of edge-sharing $FeSe_4$-tetrahedra layers separated by Na. The Fe-Se and Fe-Fe bond lengths according to the refined parameters are 2.362(4) Å and 2.676(3) Å, respectively, shrinking a lot compared over with 2.4406(4) Å and 2.7673(5) Å for $K_{0.8}Fe_2Se_2$[1]. Bond angles, however, show only slight changes. Similarly, the refinement results for $Ba_{0.8}Fe_2Se_2$ are shown in Supplementary Figure S1 and Table SI. Indexing the pattern for $KFe_2Se_2$ by the ammonothermal method yields two body-centered tetragonal cells with lattice constants $a$ = 3.755(3) Å and $c$ = 20.48(1) Å, and another with lattice constants $a$ = 3.79(2) Å and $c$ = 16.16(5) Å. Similar situation applies to $Ca_{0.5}Fe_2Se_2$, $Sr_{0.8}Fe_2Se_2$, $EuFe_2Se_2$, and $Yb_{0.8}Fe_2Se_2$, in which the main diffraction peaks can be indexed based on two body-centered tetragonal cells. This multi-phase nature reflects the inhomogeneity of these samples for the intercalation process is diffusion controlled. All the indexing results are listed in Supplementary Table SII. Examination of the intercalated compounds reveals that all the lattice constants $a$ expand, while $c$ are nearly tripled or even larger compared over $\beta$-FeSe[27]. This result along with the chemical analysis (see Table II) indicates that the metals are indeed intercalated in between FeSe layers and form new layered compounds similar to $K_{0.8}Fe_{1.7}Se_2$ by the high-temperature route in structure. But the inserting amounts are varying from metal to metal.

The magnetism of samples was measured as a function of temperature or magnetic field. Shown in Fig. 3 are the *M-T* curves for nominal $NaFe_2Se_2$ under zero field and a field of 40 Oe cooling, respectively. Over the temperature range from room temperature to the onset transition temperature, zero-field-cooling (ZFC) and field-cooling (FC) curves are quite flat and nearly temperature independent, implying little or no magnetic impurity ($\alpha$-FeSe, $Fe_7Se_8$, Fe and etc.) in the sample. The curves sharply drop in both ZFC and FC signaling a SC transition begins at about 46 K, comparable to 48.7 K observed in $K_{0.8}Fe_{1.7}Se_2$ under a pressure of 12.5 GPa[28]. To confirm the SC transition, we measured its *M-H* curves at 10 K and 60 K and *ρ-T* curve (see inset of Fig. 3). The *M-H* curve at 10 K exhibits clear magnetic hysteresis, a typical feature of type-II superconductor. The superconductive shielding fraction is estimated to be about 40 % at 10 K. A sharp transition in electrical resistance at 45 K is observed though not reaching zero resistance as the sample is cold pressed, further confirming the SC transition. Figure 4a shows the *M-T* curves for nominal $Ba_{0.8}Fe_2Se_2$, very similar to the ones for $NaFe_2Se_2$, except for the transition temperature that is lowered to 39 K. The estimated superconductive shielding fraction is about 62 % at 10 K. Shown in the inset of Fig. 4a are *M-H* curves and *ρ-T* curve of $Ba_{0.8}Fe_2Se_2$, confirming its SC transition and the type-II superconductor again. Its lower critical field ($H_{c1}$) is around 0.15 T and the estimated upper critical filed ($H_{c2}$) is about 36.5 T (see Supplementary Fig. S2). Figure 4b shows the heat capacity of cold-pressed powder $Ba_{0.8}Fe_2Se_2$. In the normal state, the electronic coefficient of heat capacity $\gamma$ is determined to be $3.47 \times 10^{-3}$ mJ g$^{-1}$ K$^{-2}$ according to the zero-temperature intercept. At lower temperatures, no apparent peak is present, but a clear deviation from the linear change at 39 K is observed, again confirming the nature of bulk SC in Ba$_{0.8}$Fe$_2$Se$_2$. For all other samples, the *M-T* and *M-H* curves can be seen in Fig. 5. In particular, we note that the Li insertion leads to SC at 44 K, in contrast to the previous report that Li intercalation into Fe(Se,Te) has no effect on SC and structure[29]. Further work is underway to clarify the structure and property of these new superconductors.

Some samples obtained by the ammonothermal method are sensitive to atmosphere. Magnetization measurements on sample KFe$_2$Se$_2$ show its SC transition temperature changes from 40 K to 30 K after one hour's exposure to air (see Supplementary Fig. S3), accompanying the disappearance of one body-centered tetragonal cell with lattice constants $a = 3.755(3)$ Å, $c = 20.48(1)$ Å shown in Supplementary Fig. S4. Others, however, are relatively stable. For example, the SC transition temperature for Ba$_{0.5}$Fe$_2$Se$_2$ shows no apparent change after several hours in air and 4 days in glove box, only a 4 K drop after annealing at 373 K for 3 days (see Supplementary Fig. S5). The powder diffraction patterns (see Supplementary Fig. S6) show that only the impurity Ba(OH)$_2$ turns into Ba(OH)$_2 \cdot$H$_2$O after air exposure and annealing, while the diffraction peaks corresponding to Ba$_{0.5}$Fe$_2$Se$_2$ show no apparent change. NH$_3$ is possible to enter in between the layers as observed in MoS$_2$[24] and C$_{60}$[30,31]. But infrared spectroscopy measurements on our samples do not show clear peak due to N-H vibrations (see Supplementary Fig. S7) and the role of NH$_3$ in inducing SC is not clear, which is inconsistent with the recent report of Li$_x$(NH$_2$)$_y$(NH$_3$)$_{1-y}$Fe$_2$Se$_2$ in which lithium ions, lithium amide and ammonia are intercalated into FeSe layers[26]. The SC of samples only soaking in LA without addition of metal has not been confirmed above 10 K.

**Discussion**

The refined structures manifest that Na and Ba are intercalated in between FeSe layers. But it is hard to understand why their lattice parameters $c$ are much elongated compared with the $c$ for $K_{0.8}Fe_2Se_2$[1]. This is, in particular, true for $NaFe_2Se_2$ as Na has a smaller ionic radius than K has. One plausible explanation is that in between FeSe layers there exist other atoms or atomic groups such as $NH_3$, $NH_2^-$ apart from Na and Ba, though infrared spectra do not show the vibration peak due to N-H bond. Their accurate structures, however, need to be confirmed by other experimental means in future. In addition, we note that the refined Ba occupancy is quite low as 0.212(3), less than half of that for Na in $NaFe_2Se_2$. This is reasonable from the point of charge transfer view since its normal valence is +2.

Hence, SC at 30~46 K in these intercalated iron selenides is believed to be induced by high electron doping from the metals intercalated to FeSe layers, which has been previously proved in $Am_{0.8}Fe_2Se_2$[1-7] (Am = K, Rb, Cs, and Tl). Further work is presently underway to understand the differences between the structure and onset transition temperatures for the Li, Na, Ba, Sr, Ca, Yb, and Eu intercalates. Although here we have focused on the intercalation of alkali metals, alkaline earths, and some rare earths, the approach to intercalate other spacer layers and to synthesize from solvents other than LA should also be possible to adopt in future. In particular, compounds with unit cells large enough are promising parent targets to obtain new superconductors with possible higher $T_c$.

In summary, we have obtained a series of new superconductors through the ammonothermal method at room temperature. A variety of metals are verified to be intercalated in between FeSe layers to induce the SC with various enhanced transition temperatures. The results presented here demonstrate that superconductors are rich in M-FeSe system induced by doping.

## Methods

In this work, we first synthesized $\beta$-FeSe powders following the method described in ref. 32. Alkali metals, alkaline earths, Yb, Eu, and $\beta$-FeSe powder in molar ratios A/$\beta$-FeSe ≈ (0.5~1):1 were loaded in a 23 ml or 50 ml autoclave which was placed in a bath of liquid nitrogen for 2-3 minutes and then LA (99.999%) was slowly filled up to the 1/3~1/2 volume of the autoclave with the concentrations of metals solved in LA reaching about 0.1~0.3 atomic %. All the conducts were performed under Ar atmosphere in a glove box to prevent air and water contamination. After sealed, the autoclaves were taken out and kept at room temperature for 2-17 days. During the process, shaking was needed to facilitate the reaction and to improve the homogeneity of the products. Finally, the samples were rinsed by using fresh LA to eliminate soluble impurities. All the samples' nominal compositions, masses of starting materials, and synthetic conditions are listed in Table II. Cautionary note: caution should be taken to avoid frostbite by LA.

The samples were characterized by powder X-ray diffraction using a PANalytical X'pert PRO diffractometer with Cu $K\alpha$ radiation. Indexing was performed with DICVOL06[33]. Rietveld refinements of the diffraction data were performed with the FULLPROF package[34]. The magnetic and transport properties were characterized using a vibrating sample magnetometer (VSM, Quantum Design) and the physical property measurement system (PPMS, Quantum Design), respectively. The magnetic measurements were carried out in *dc* field of 40 Oe in the temperature range 5-300 K after cooling in zero field and in the measuring field. The electrical resistances were measured by using the standard four-probe method based on samples cold-pressed at a uniaxial stress of 400 kg cm$^{-2}$. The low temperature heat capacity was measured on the cold-pressed powder sample by thermal relaxation method using PPMS.

neutron powder diffraction. *Physica B* **192**, 55-69 (1993).

**Acknowledgements**

This work was partly supported by the National Science Foundation of China under Grant Nos. 90922037 and 51072226, and by the Chinese Academy of Sciences.

**Author contributions**

T.P.Y. did most of the synthesis and characterizations. X.L.C. designed the experimental scheme. X.L.C and G.W. analysed the data and wrote the paper. S.F.J. did the structure analysis. T.T.Z., X.F.L., H.Z., and W.Y.W. helped with the experiment.

**Additional information**

**Supplementary Information** accompanies this paper at http://www.nature.com/scientificreports.

**Competing financial interest:** The authors declare no competing financial interests.

| Table I | Crystallographic data and Rietveld refinement data for $NaFe_2Se_2$ |
|---|---|
| Temperature (K) | 297 |
| Space group | $I4/mmm$ |
| Fw | 292.6 |
| $a$ (Å) | 3.7846(4) |
| $c$ (Å) | 17.432(1) |
| $V$ (Å$^3$) | 249.69(4) |
| Z | 2 |
| $R_p$ | 2.18% |
| $R_{wp}$ | 3.05% |
| $R_{exp}$ | 1.77% |
| GoF-index | 1.7 |
| Atomic parameters | |
| Na | 2a (0, 0, 0) |
| Occ (Na) | 0.67(2) |
| Fe | 4d (0, 0.5, 0.25) |
| Se | 4e (0, 0, z) |
|  | z = 0.3311(4) |
| Bond length (Å) | |
| Na-Se | 3.979 (5) ×8 |
| Fe-Se | 2.362 (4) ×4 |
| Fe-Fe | 2.676 (3) ×4 |
| Bond angles (deg.) | 111.0 (1) ×4 |
|  | 106.5 (2) ×2 |

| Table II | Nominal compositions, masses of starting materials, and synthetic conditions for $A_xFe_2Se_2$ (A = Li, Na, K, Ba, Sr, Ca, Yb, and Eu) | | | | | | |
|---|---|---|---|---|---|---|---|
| Nominal composition | Masses used | | | Reaction duration (days) | $T_c^{onset}$ (K) | Superconductive shielding fraction (10 K) | Composition characterized by ICP-AES |
| | Metal (g) | FeSe (g) | LA (ml) | | | | |
| $LiFe_2Se_2$ | 0.0077 | 0.3001 | 15 | 8 | 44 | 32% | - |
| $Na_{0.5}Fe_2Se_2$ | 0.0175 | 0.4001 | 15 | 3 | 45 | ~34% | $Na_{0.42}Fe_2Se_{1.88}$ |
| $NaFe_2Se_2$ | 0.0272 | 0.3002 | 15 | 6 | 46 | 50% | $Na_{0.61}Fe_2Se_{1.9}$ |
| $NaFe_2Se_2$ | 0.0258 | 0.3002 | 15 | 3 | 46 | 40% | $Na_{0.69}Fe_2Se_{1.96}$ |
| $NaFe_2Se_2$ | 0.0259 | 0.3006 | 15 | 4 | 46 | 27% | $Na_{0.8}Fe_2Se_2$ |
| $K_{0.5}Fe_2Se_2$‡ | 0.0220 | 0.3000 | 10 | 13 | 29 | 1.5% | - |
| $KFe_2Se_2$§ | 0.0290 | 0.3999 | 15 | 2.5 | 40 | 11% | - |
| $KFe_2Se_2$§ | 0.0290 | 0.3999 | 15 | 2.5 | 30 | 2.5% | - |
| $Ca_{0.5}Fe_2Se_2$ | 0.0226 | 0.3000 | 10 | 2 | ~40 | 4% | - |
| $Sr_{0.5}Fe_2Se_2$ | 0.0491 | 0.3010 | 15 | 8 | ~35 | 10% | - |
| $Sr_{0.8}Fe_2Se_2$ | 0.0814 | 0.3131 | 15 | 13 | 38 | 60% | - |
| $Ba_{0.5}Fe_2Se_2$ | 0.0509 | 0.2000 | 10 | 8 | 40 | 25% | $Ba_{0.48}Fe_2Se_{1.16}$ |
| $Ba_{0.8}Fe_2Se_2$ | 0.1219 | 0.3000 | 10 | 4 | 39 | 62% | $Ba_{0.64}Fe_2Se_{1.62}$ |
| $EuFe_2Se_2$ | 0.0843 | 0.3001 | 15 | 10 | 40 | 8% | - |
| $Yb_{0.5}Fe_2Se_2$ | 0.0963 | 0.3000 | 15 | 10 | 42 | 7% | - |
| $Yb_{0.8}Fe_2Se_2$ | 0.1365 | 0.2658 | 10 | 10 | 42 | 5% | - |
| FeSe (soaked in LA)‡ | - | 0.2000 | 10 | 17 | 8.5◊ | - | - |

The superconductive shielding fraction was estimated at 10 K from ZFC curve. ICP-AES, inductively coupled plasma atomic emission spectrometer.

‡ Samples synthesized without shaking.

§ Same sample measured before and after 1 hour's exposure to air (see Supplementary Fig. S3).

◊ SC due to β-FeSe.

**Figure 1 Powder X-ray diffraction patterns for samples measured at 297 K, Cu Kα radiation. a,** $\beta$-FeSe; **b,** $K_{0.8}Fe_{1.7}Se_2$ (by high-temperature route); **c,** Nominal $LiFe_2Se_2$; **d,** Nominal $NaFe_2Se_2$; **e,** Nominal $KFe_2Se_2$ (background corrected), peaks marked by '▽' are due to unknown phase; **f,** Nominal $Ca_{0.5}Fe_2Se_2$; **g,** Nominal $Sr_{0.8}Fe_2Se_2$; **h,** Nominal $Ba_{0.8}Fe_2Se_2$; **i,** Nominal $EuFe_2Se_2$; **j,** Nominal $Yb_{0.8}Fe_2Se_2$. Peaks marked by '*' are due to residual $\beta$-FeSe.

**Figure 2 Powder X-ray diffraction pattern and Rietveld refinement profile for nominal $NaFe_2Se_2$ at 297 K.** Vertical bars (│) indicate the positions of the Bragg peaks. The bottom trace depicts the difference between the experimental and calculated intensity values. The inset shows the crystal structure of $NaFe_2Se_2$.

**Figure 3 Magnetization and electrical resistance of nominal $NaFe_2Se_2$.** The left inset shows the magnetic hysteresises of nominal $NaFe_2Se_2$ measured at 10 K and 60 K in the range -25 kOe < H < 25 kOe, respectively. The right inset shows the temperature dependence of the electrical resistance of cold-pressed powder $NaFe_2Se_2$.

**Figure 4 Magnetization, electrical resistance, and heat capacity of nominal $Ba_{0.8}Fe_2Se_2$. a,** The magnetization of nominal $Ba_{0.8}Fe_2Se_2$ as a function of temperature. The left inset shows the magnetic hysteresises of nominal $Ba_{0.8}Fe_2Se_2$ measured at 10 K and 60 K in the range -25 kOe < $H$ < 25 kOe, respectively. The right inset shows the temperature dependence of the electrical resistance of cold-pressed powder

Ba$_{0.8}$Fe$_2$Se$_2$. **b,** Low temperature heat capacity of cold-pressed powder Ba$_{0.8}$Fe$_2$Se$_2$. The red dotted line is the curve fitting of phonon contribution to the heat capacity.

**Figure 5 Magnetizations of nominal A$_x$Fe$_2$Se$_2$ (A = Li, K, Ca, Sr, Eu, and Yb). a.** The magnetization of nominal LiFe$_2$Se$_2$ as a function of temperature. The inset shows the magnetic hysteresises of nominal LiFe$_2$Se$_2$ measured at 10 K and 40 K. **b.** The magnetization of nominal KFe$_2$Se$_2$ as a function of temperature. The inset shows the magnetic hysteresises of nominal KFe$_2$Se$_2$ measured at 10 K and 35 K. **c.** The magnetization of nominal Ca$_{0.5}$Fe$_2$Se$_2$ as a function of temperature. The inset shows the magnetic hysteresises of nominal Ca$_{0.5}$Fe$_2$Se$_2$ measured at 10 K and 50 K. **d.** The magnetization of nominal Sr$_{0.8}$Fe$_2$Se$_2$ as a function of temperature. The inset shows the magnetic hysteresises of nominal Sr$_{0.8}$Fe$_2$Se$_2$ measured at 10 K and 50 K. **e.** The magnetization of nominal EuFe$_2$Se$_2$ as a function of temperature. The inset shows the magnetic hysteresises of nominal EuFe$_2$Se$_2$ measured at 10 K and 50 K. **f.** The magnetization of nominal Yb$_{0.8}$Fe$_2$Se$_2$ as a function of temperature. The inset shows the magnetic hysteresises of nominal Yb$_{0.8}$Fe$_2$Se$_2$ measured at 10 K and 60 K.

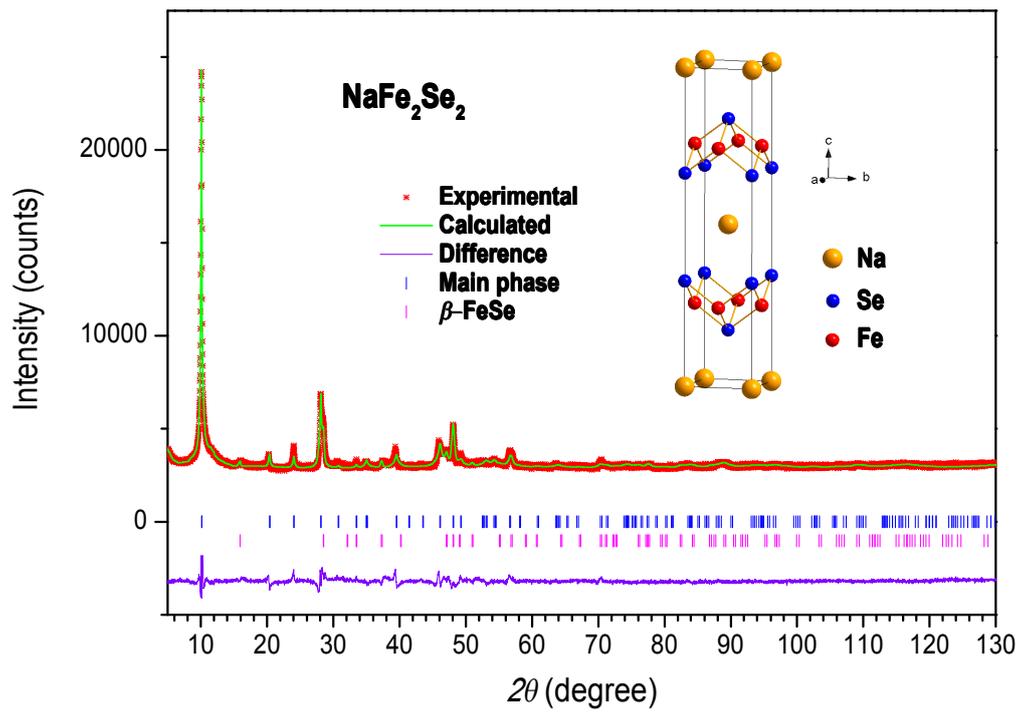

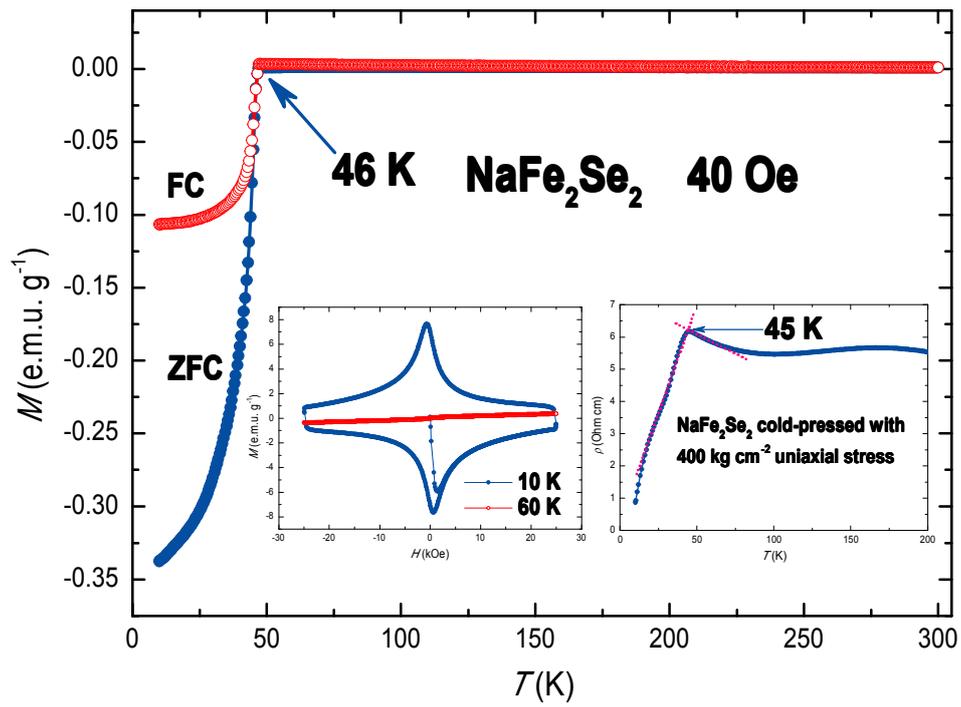

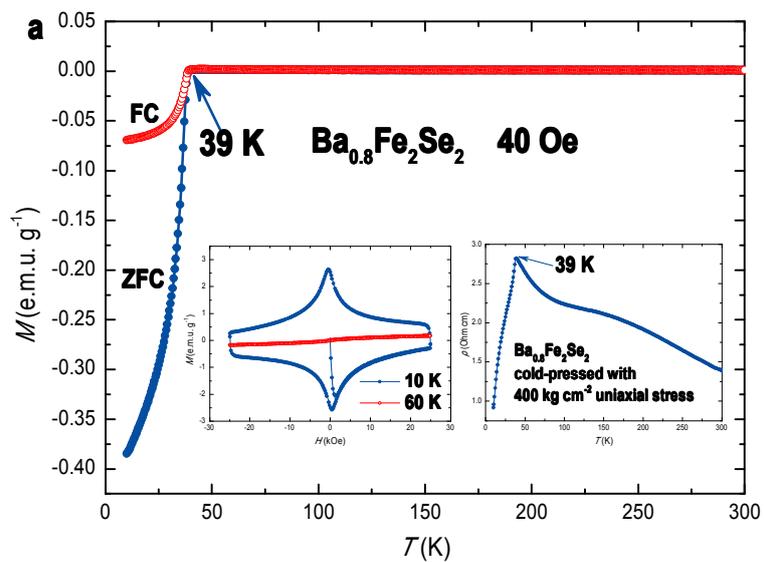

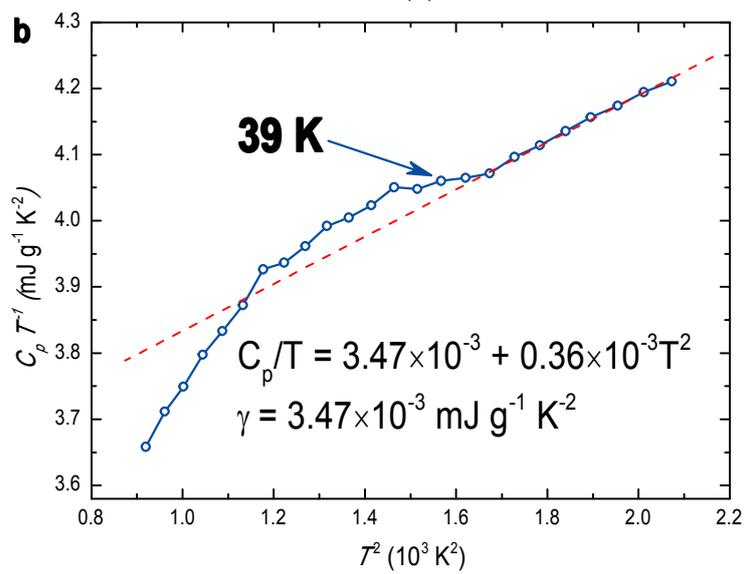

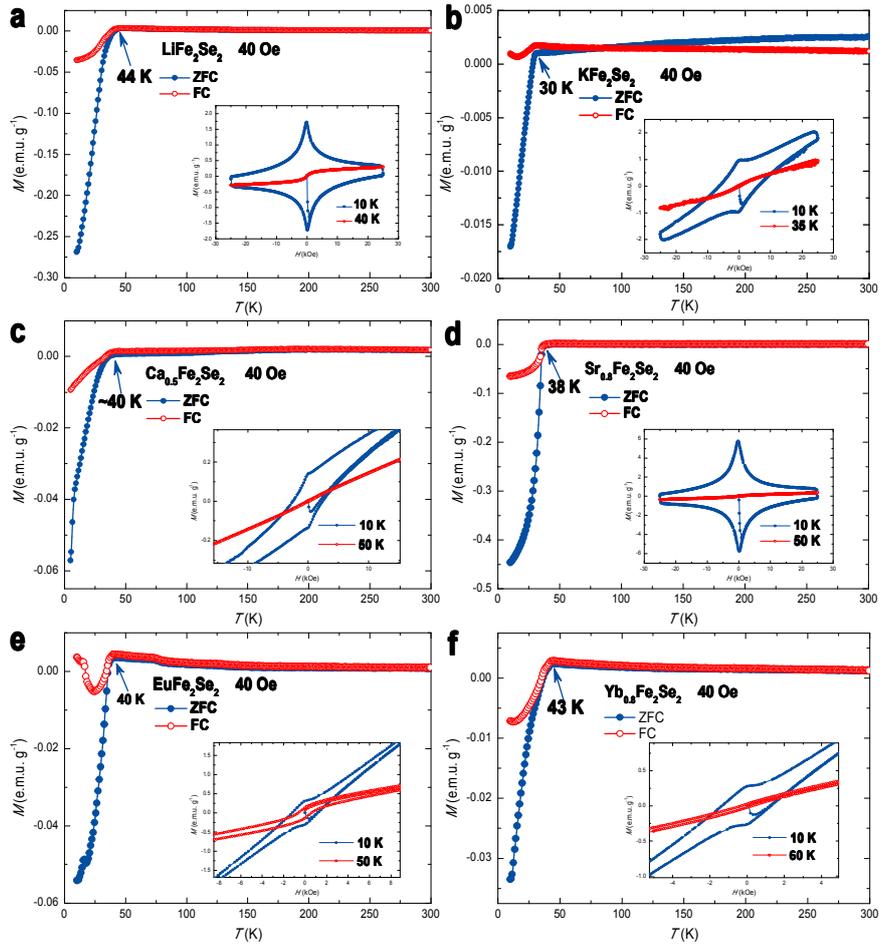